\documentstyle[12pt]{article}

\newcommand{\bg}[1]{\mbox{\boldmath $#1$}}

\title{
~\vspace{-3cm}
\begin{center}\begin{Large}\begin{bf}
 UNIVERSIT\'E DE GEN\`EVE\\[-0.4cm]
  \end{bf}\end{Large}\smallskip\begin{small}
 SCHOLA GENEVENSIS MDLIX
  \end{small}\end{center}\goodbreak\begin{center}
\bigskip\vbox
{\vskip 3truecm\noindent
\vskip 2truecm\noindent}
\end{center}
{\large\bf 
CORRECTIONS TO THE PAGELS-STOKAR FORMULA FOR $f_{\pi}$
\thanks{Partially 
supported by the Swiss National Foundation}}  
\vspace{1cm}}
\author{\large
A.Barducci, R.Casalbuoni, M.Modugno, G.Pettini\\
{\small Dipartimento di Fisica, Univ. di Firenze and
I.N.F.N., Sezione di Firenze }\\[0.5cm]
R.Gatto\\
{\small D\'epartement de Physique Th\'eorique, Univ. de Gen\`eve}
\vspace{1cm}}
\date{UGVA-DPT 1997/03-972\\
Firenze Preprint - DFF-273/03/1997}

\begin{document}

\maketitle

\thispagestyle{empty}
\newpage 

\vspace{1cm}
\begin{abstract}
\indent\noindent

Within the composite operator formalism we derive a formula for the pion decay
constant $f_{\pi}$, as defined directly from the residue at the pion pole of 
the meson propagator, rather than from the matrix element of the axial current.
The calculation is performed under some simplifying assumptions, and we verify 
the complete consistency with soft-pion results, in 
particular with the Adler-Dashen relation.
 The formula one obtains for (the pole-defined)
$f_{\pi}^2$ differs from the previous Pagels-Stokar expression by an additive 
term, and it still provides $f_{\pi}^2$ in terms of the quark self-energy. We
make some numerical estimates leading to $(30 \div 40)\%$ deviation 
for $f_{\pi}^2$ with respect to the Pagels-Stokar formula.

\end{abstract}
\newpage

\section{Introduction}
\indent\noindent

In quantum chromodynamics the pion, as well-known, is a pseudo 
Nambu-Goldstone boson associated with the spontaneous breaking of chiral 
symmetry. Its decay constant $f_{\pi}$ plays a key dynamical role in the chiral 
symmetry breaking mechanism of QCD, and analogous quantities appear in other
theories which use similar paradigms, such as electroweak symmetry breaking 
through a new strong sector. In 1979 Pagels and Stokar proposed an approximate
expression for the calculation of $f_{\pi}$ \cite{pagels}.
Their derivation used a sum rule due to Jackiw and Johnson \cite{jackiw}, 
plus additional assumptions within the 
so-called dynamical perturbation theory, and allowed for an
approximate expression for $f_{\pi}$, as defined from the matrix element of the
axial current, in terms of the quark self-energy. The
formula is of great utility in QCD and also in theories derived from the old 
technicolour concept. 

There has been a vast literature on the 
Pagels-Stokar (PS) formula with the general conclusion that it leads to a 
sensible result for $f_{\pi}$, within uncertainties not easily controllable
in view of the theoretical approximations present in the 
derivation \cite{miranski}.
The Pagels-Stokar expression for the pion decay constant is

\begin{eqnarray}
f_{PS}^2&=&{d(\underline{r})\over(2\pi)^2}\int_0^\infty d k^2~ k^2~
{\Sigma_{0}^2(k^2)-{1\over 2} k^2 \Sigma_{0}(k^2)\displaystyle{
d \Sigma_{0}(k^2)\over d k^2}\over
\left[\displaystyle k^2+\Sigma_{0}^2(k^2)\right]^2}\\
&&\nonumber 
\end{eqnarray} 
where $d(\underline{r})$ is the dimension of the quark colour representation 
($d(\underline{r})=3$ in QCD)
and $\Sigma_{0}$ is the dynamical quark self-energy in the chiral limit.

We shall present below a new formula for $f_{\pi}$, which shares with the PS
formula the advantage of only depending on the self-energy $\Sigma_{0}$, and
that we derive within the composite operator formalism developed 
in ref. \cite{cjt} and as modified in ref. \cite{bcd}. Within such schemes one 
could think of two
different calculations of the pion decay constant. One can evaluate the
coupling of the pseudo-Goldstone to the axial-vector current. Alternatively,
one can directly evaluate the residue at the pion pole of the meson 
propagator. These procedures correspond to different definitions, but 
only the second one agrees with the current-algebra and
soft-pion results \cite{genrev}
 (in particular the Adler-Dashen relation). We shall follow 
the second procedure, that is calculating the residue at the pion pole.
Within the composite operator formalism, 
for a first approximate understanding, we discuss here the so-called
``rigid case'', in which the presumably small logarithmic corrections 
coming from the 
renormalization-group analysis are neglected. For the asymptotic behaviour
of the self-energy function we choose the one dictated by the operator 
product expansion (OPE) \cite{miranski} up to a logarithm which we neglect.
The motivation for such approximate study is mainly that in this way
it is possible to derive analitically the complete expression for the
effective action at two fermion loops. Furthermore a phenomenological
analysis we have previously performed of the pseudoscalar masses \cite{bcd}
in this contest did not show any major inadequacy of such a treatment.
Our expression for $f_{\pi}^2$ is  

\begin{eqnarray}
f_{\pi}^2 &=& {d(\underline{r})\over(2\pi)^2}
\int_0^\infty d k^2~\left[
k^2~{\Sigma_{0}^2(k^2)-{1\over 2} k^2 \Sigma_{0}(k^2)
{\displaystyle {d \Sigma_{0}(k^2)\over{d k^2}}}\over
\big[k^2+\Sigma_{0}^2(k^2)\big]^2}
\right.\nonumber\\
&&\qquad\qquad+\left.{\displaystyle k^6 
\left({d\Sigma_{0}(k^2)\over d k^2}\right)^2
- k^4\Sigma_{0}^2(k^2)\left({d \Sigma_{0}(k^2)\over d k^2 }\right)^2 - k^4
\Sigma_{0}(k^2){d \Sigma_{0}(k^2) \over d k^2 }\over 2
{\big[k^2+\Sigma_{0}^2(k^2)\big]^2}
}\right]
\end{eqnarray}

The first term in Eq. (2) is $f_{PS}^2$ of Eq. (1). By writing 
$f_{\pi}^2$ = $f_{PS}^2$ $(1+{\delta}^2)$, to get an evaluation of $\delta$
we can go back to two different Ansatz for $\Sigma_{0}$ we had used 
in the past to study low energy QCD \cite{bcd,giulio}.
We find for $\delta^2$ values such as 0.35 and 0.37, suggesting
that the correction to the Pagels-Stokar expression is presumably large
$(30 \div 40)\%$ and presumably rather insensitive 
to the modification of the self-energy shape, provided the ultra-violet 
behaviour in $k^2$ is roughly maintained.

\section{The effective action}
\indent\noindent

We start from the effective Euclidean action for the composite operator 
formalism
\begin{equation}
\Gamma( {\bf\Sigma} ) = - {\rm Tr}\ln\left[{\bf S}_0^{-1} + {\delta
\Gamma_2 \over\delta{\bf S}}\right] + {\rm Tr}\left[ {\delta \Gamma_2
\over\delta{\bf S} }{\bf S}\right] -{\bf \Gamma}_2({\bf S})
+ counterterms
\end{equation}
where ${\bf S}^{-1}_0= (i\hat{p}- {\bf m})$, {\bf m} is the bare 
quark mass matrix, ${\bf \Gamma}_2({\bf S})$ 
is the sum of all two-particle irreducible vacuum diagrams with
fermionic propagator ${\bf S}$ and ${\bf \Sigma} = 
-\delta{\bf \Gamma}_2/\delta{\bf S}$. Eq. (3) is the modification 
of the effective  action of Cornwall, Jackiw and Tomboulis
\cite{cjt} which was introduced in ref. \cite{bcd} to account for 
the correct stability properties of the theory. 
At two-loop level ${\bf \Gamma}_2=
{1\over2}{\rm Tr}({\bf S}\Delta{\bf S})$, where $\Delta$ is the gauge 
boson propagator, so that ${\bf \Sigma} = - \Delta{\bf S}$, 
${\rm Tr}\left[{\delta{\bf \Gamma}_2/\delta{\bf S}}\right] = 
2{\bf \Gamma}_2$, and one can rewrite Eq. (3) in terms of ${\bf \Sigma}$
\begin{eqnarray}
\Gamma( {\bf\Sigma} ) &=& - {\rm Tr}\ln\left[{\bf S}_0^{-1} -{\bf \Sigma}
\right] + {\bf \Gamma}_2({\bf \Sigma}) + counterterms\nonumber\\
&=&- {\rm Tr}\ln\left[{\bf S}_0^{-1} -{\bf \Sigma}
\right] + {1\over 2}{\rm Tr}\left({\bf \Sigma}\Delta^{-1}{\bf \Sigma}
\right) + counterterms
\end{eqnarray}

Here the variable ${\bf \Sigma}$ plays the role of a dynamical
variable. At the minimum of the functional action, that is when
the Schwinger-Dyson equation is satisfied, ${\bf \Sigma}$ is nothing but 
the fermion self-energy .
A parametrization for ${\bf \Sigma}$, employed in \cite{bcd}, was
\begin{equation}
 {\bf\Sigma}=({\bf s} + i \gamma_5{\bf p})f(k)\equiv {\bf\Sigma}_s
+i\gamma_5 {\bf\Sigma}_p
\end{equation}
with a suitable Ansatz for $f(k)$, and with {\bf s} and {\bf p} scalar
and pseudoscalar constant fields respectively.

\section{The effective potential}
\indent\noindent

The effective potential one obtains from Eq. (4) (see ref. \cite{bcd}) is
\begin{eqnarray}
V={\Gamma\over\Omega}&=& -{8\pi^2 d(\underline{r})\over 3 C_2(\underline{r})
g^2}
\int{d^4 k\over(2\pi)^4}{\rm tr}\left[{\bf \Sigma}_s\Box_k{\bf \Sigma}_s
+{\bf \Sigma}_p\Box_k{\bf \Sigma}_p\right]-\nonumber\\
&&- d(\underline{r})~{\rm Tr}\ln\left[i\hat{k} - \left({\bf m}+{\bf \Sigma}_s
\right)-i\gamma_5{\bf \Sigma}_p\right] + \delta Z~{\rm tr}({\bf m}{\bf s})
\end{eqnarray}
where $C_2(\underline{r})$ is the quadratic Casimir of the fermion colour 
representation (for $SU(3)_c$ $C_2 = 4/3$) and 
${\bf \Sigma}_s = \lambda_{\alpha} s_{\alpha} f(k)/ \sqrt{2}$, 
${\bf \Sigma}_p = \lambda_{\alpha} p_{\alpha} f(k)/ \sqrt{2}$, 
${\bf m} = \lambda_{\alpha}m_{\alpha}/ \sqrt{2}$ ($\alpha=0,\cdots,8$,
$\lambda_0 = \sqrt{2/3}$, $\lambda_i =$ Gell-Mann matrices, $i=1,\cdots,8$).
Furthermore $\delta Z$ has a divergent piece
to compensate the leading divergence proportional to ${\rm tr}({\bf m}{\bf s})$
in the logarithmic term.
For a quark of mass $m$ the effective potential is
\begin{eqnarray}
V(s, p, m) &=& -{ d(\underline{r}) c}
\int{d^4 k\over(2\pi)^4}\left[{\Sigma}_s\Box_k{ \Sigma}_s
+{ \Sigma}_p\Box_k{ \Sigma}_p\right]-\nonumber\\
&&- 2d(\underline{r})\int{d^4 k\over(2\pi)^4}
\ln\left[k^2 + \left( m+{\Sigma}_s\right)^2 + {\Sigma}_p^2\right] 
+ \delta Z m s
\end{eqnarray}
where we have defined $c=2\pi^2/g^2$.
In ref. \cite{bcd} after fixing $\delta Z$ so as to cancel the leading 
divergence proportional to $m s$ in the logarithm, we had imposed the
normalization condition
\begin{equation}
\lim_{ m\rightarrow0}{1\over m}\left.{\partial V\over
\partial\langle\bar{\psi}\psi\rangle}\right|_{extr}=1
\end{equation}
or, with $\langle\bar{\psi}\psi\rangle = 
\left(d(\underline{r})M^3/2\pi^2\right) c \bar{s}$
\begin{equation}
\lim_{ m\rightarrow0}{1\over m}\left.{\partial V\over
\partial s}\right|_{extr}={d(\underline{r})M^3\over2\pi^2}c
\end{equation}
where $M$ is a momentum scale for the self energy and $\bar{s}$ is the 
extremum of the effective potential.
The extrema of the effective potential in the massless case $m=0$ depend only
on $c$. Therefore Eq. (9) becomes, in this case, an 
equation for $c$ and M is left
undetermined. This is nothing but the usual dimensional transmutation.
The numerical values for $c$ and $s_0$, the minimum of the effective potential
in the massless case, are obtained once one has fixed the Ansatz for $f(k)$.  

With $\Sigma_0(k) = s_0 f(k)$ we shall write
in general
\begin{equation}
\delta Z = d(\underline{r})\left[{M^3\over2\pi^2}c +
{4\over s_0}\int{d^4 k\over(2\pi)^4}{
{\Sigma}_{0s}\over k^2+ {\Sigma}_{0s}^2}\right]
\end{equation}
The gap equation, from $\displaystyle{d V\over d s}=0$, is 
\begin{equation}
{d(\underline{r})\over \bar{s}}\left[-2c \int{d^4 k\over(2\pi)^4}
\bar{\Sigma}_s\Box_k\bar{\Sigma}_s - 4\int{d^4 k\over(2\pi)^4}
{(m + \bar{\Sigma}_s)\bar{\Sigma}_s\over k^2+ (m+\bar{\Sigma}_s)^2}\right]
+m\delta Z = 0
\end{equation}
where $\bar{\Sigma}_s=\bar{s}f(k)$ and $\bar{s}$ is
the value at the minimum in the presence of the bare mass.

Let us now turn to the effective action. The fields {\bf s}
and {\bf p} depend in this case on the space coordinates and we shall 
use the Weyl symmetrization prescription
\begin{equation}
 {\bf\Sigma}=({\bf s} + i \gamma_5{\bf p})f(k)\rightarrow
{1\over2}\left[
{\bf s}({\bf x}) + i\gamma_5 {\bf p}({\bf x}),f({\bf k})
\right]_{+}
\end{equation}
We are interested in oscillations around the minimum of the 
effective potential, so we introduce
\begin{eqnarray}
&&{\bg\chi}({\bf x})={\bf s}({\bf x})-\bar{\bf s},\qquad
{\bg\pi}({\bf x})={\bf p}({\bf x})-\bar{\bf p}\equiv{\bf p}({\bf x})
\nonumber\\
&&{\bf v}({\bf x})={\bg\chi}({\bf x})+i\gamma_5{\bg \pi}({\bf x})
\nonumber\\
&&\bar{\bf S }(k)=i\hat{k}- \left( ({\bf m} +\bar{\bf \Sigma}_s(k)\right)
\end{eqnarray}
The {\em Tr ln} term in Eq. (4), which we denote as  $\Gamma_{log}$, becomes
\begin{equation}
\Gamma_{log}=-{\rm Tr}\ln \left[i\hat{k} - {\bf m} -{\bf \Sigma}\right]
=-{\rm Tr}\ln \left[\bar{\bf S }^{-1} -{1\over2}\left[
{\bf v}({\bf x}),f({\bf k})\right]_{+}\right]
\end{equation}
As we are interested in the 2-points function, we expand to second
order in ${\bf v}({\bf x})$ and after some calculation we obtain
for the Fourier transform of $\Gamma_{log}$ in Eq. (14)
\begin{eqnarray}
\Gamma_{log}&=&d(\underline{r})\Big\{
-2\Omega\int{d^4k\over(2\pi)^4}
{\rm tr}\ln\left[
k^2 + \left({\bf m} +\bar{\bf \Sigma}_s(k)\right)^2
\right] +\int{d^4k\over(2\pi)^4}f(k){\rm tr}
\left[\bar{\bf S}(k){\bg\chi}(0)\right]
\nonumber\\
&&+
{1\over2}\int{d^4k\over(2\pi)^4}\int{d^4q\over(2\pi)^4}
{\rm tr}\left[
\bar{\bf S}(k)i\gamma_5 V(k, k+q){\bg\pi}(-q)\bar{\bf S}(k+q)
i\gamma_5 V(k+q, k){\bg\pi}(q)\right. \nonumber\\
&&+\Big. \Big. (pseudoscalar\leftrightarrow scalar,
i\gamma_5\leftrightarrow {\bf 1})\Big]+ \cdots\Big\}
\end{eqnarray}
where
\begin{equation}
V(k_1,k_2)\equiv {1\over2}\left[f(k_1) + f(k_2)\right]
\end{equation}

\newpage

For $\Gamma_2$ we obtain (working in Landau's gauge)
\begin{eqnarray}
\Gamma_2&=& -{d(\underline{r})\; c}\;\Omega
\int{d^4 k\over(2\pi)^4}{\rm tr}\left[\bar{\bf \Sigma}_s
\Box_k\bar{\bf \Sigma}_s\right]\nonumber\\
&& -{2d(\underline{r}) c}
\left(\int{d^4 k\over(2\pi)^4}f(k)\Box_k f(k)\right){\rm tr}
(\bar{\bf s}{\bg\chi}(0))\nonumber\\
&& +\int{d^4 q\over(2\pi)^4}{\rm tr}
\Big\{
-{d(\underline{r}) c}{\bg \pi}(-q)
\left( \int{d^4 k\over(2\pi)^4}f(k)\Box_k f(k) \right)
{\bg \pi}(q)
\nonumber\\
&&+ (pseudoscalar \leftrightarrow scalar)\Big\}
\end{eqnarray}

For the counterterm one has
\begin{equation}
\Gamma_{ct}={\rm tr}({\bf m}\bar{\bf s})\Omega\delta Z +
 {\rm tr}({\bf m}{\bg \chi}(0))\delta Z
\end{equation}

\section{The improved expression for $f_{\pi}$}
\indent\noindent

We note that each term in the effective action consists of a constant,
a linear and a quadratic term in the fields. The constant term
gives back the original potential Eq. (6) at the minimum.
Such a term controls the normalization. The linear term vanishes by virtue
of the gap equation, Eq. (11). The quadratic term stands up for the effective 
action up to the second order in the fields. In space-time coordinates
\begin{eqnarray}
\Gamma&=&\int{d^4 x}{d^4 y}\int{d^4 q\over(2\pi)^4}
{\rm e}^{\displaystyle -iq(x - y)}\pi_{\alpha}(x)\cdot\nonumber\\
&&\cdot{\rm tr}\left\{
 -{d(\underline{r}) c}
{\lambda_{\alpha}\over\sqrt{2}}
\int{d^4 k\over(2\pi)^4}f(k)
\Box_k f(k) {\lambda_{\beta}\over\sqrt{2}}+\right.\nonumber\\
&&+\left.
{1\over2}d(\underline{r})\int{d^4k\over(2\pi)^4}
\left[
\bar{\bf S}(k)i\gamma_5{\lambda_{\alpha}\over\sqrt{2}}
 V(k, k+q)\bar{\bf S}(k+q)
i\gamma_5{\lambda_{\beta}\over\sqrt{2}}
 V(k+q, k)\right]\right\} \nonumber\\
&&\cdot\pi_{\beta}(y) +\left. (pseudoscalar\leftrightarrow scalar,
i\gamma_5\leftrightarrow {\bf 1}\right)+ \cdots
\end{eqnarray}

From
\begin{equation}
G^{-1}_{\alpha\beta}(x-y)={\delta^2\Gamma\over\delta\pi_{\alpha}(x)
\delta\pi_{\beta}(y)}
\end{equation}

one finds for the Fourier transform of $G^{-1}_{\alpha\beta}(x-y)$
\begin{eqnarray}
G^{-1}_{\alpha\beta}(q^2)&=&{\rm tr}\left\{
 -{2 d(\underline{r}) c}{\lambda_{\alpha}\over\sqrt{2}}
\int{d^4 k\over(2\pi)^4} f(k)
\Box_k f(k) {\lambda_{\beta}\over\sqrt{2}}+\right.\nonumber\\
&&+\left.
d(\underline{r})\int{d^4k\over(2\pi)^4}
\left[
\bar{\bf S}(k)i\gamma_5{\lambda_{\alpha}\over\sqrt{2}}
 V(k, k+q)\bar{\bf S}(k+q)
i\gamma_5{\lambda_{\beta}\over\sqrt{2}}
 V(k+q, k)\right]\right\}
\end{eqnarray}

By using the gap equation to eliminate $c$ we get, for a quark of mass $m$
\begin{eqnarray}
G^{-1}_{\alpha\beta}(q^2)&=&
d(\underline{r})\int{d^4k\over(2\pi)^4}
{\rm tr}\left[
\bar{S}(k)i\gamma_5 V(k, k+q)\bar{S}(k+q)
i\gamma_5 V(k+q, k)\right]\nonumber\\
&&+{4~d(\underline{r})\over\bar{s}^2}\int{d^4 k\over(2\pi)^4}
{(m + \bar{\Sigma}_s)\bar{\Sigma}_s\over 
k^2+ (m+\bar{\Sigma}_s)^2}
-{m\over\bar{s}}\delta Z
\end{eqnarray}

We can eliminate $\delta Z$ by using the normalization condition, see 
Eq. (10), and the relation
$\langle\bar{\psi}\psi\rangle = 
\left(d(\underline{r})M^3/2\pi^2\right) c \bar{s}$
to obtain
\begin{eqnarray}
G^{-1}(q^2)&=&-{m\over\bar{s}^2}\langle\bar{\psi}\psi\rangle
+ d(\underline{r})\left\{-{4 m\over\bar{s}s_0}
\int{d^4 k\over(2\pi)^4}
{{\Sigma}_{0}(k)\over k^2+{\Sigma}_{0}^2(k)}\right.\nonumber\\
&&+{4\over\bar{s}^2}\int{d^4 k\over(2\pi)^4}
{(m + \bar{\Sigma}_s)\bar{\Sigma}_s\over k^2+ (m+\bar{\Sigma}_s)^2}
\nonumber\\
&&+\left.\int{d^4k\over(2\pi)^4}
{\rm tr}\left[
\bar{S}(k)i\gamma_5V(k, k+q)\bar{S}(k+q)
i\gamma_5V(k+q, k)\right]\right\}
\end{eqnarray}
                                                     
Note that the second and third terms in Eq. (23) regularize each other
in the ultraviolet.

In the limit of small quark masses, expanding in $q^{\mu}$, we find
\begin{equation}
G^{-1}(q^2)\equiv \left({F\over\sqrt{2}s_0}\right)^2\cdot
\left(q^2 - {2m\over F^2}\langle\bar{\psi}\psi\rangle_0\right)
\end{equation}

with
\begin{eqnarray}
F^2 &=& {d(\underline{r})\over(2\pi)^2}
\int_0^\infty d k^2~\left[
k^2~{\Sigma_{0}^2(k^2)-{1\over 2} k^2 \Sigma_{0}(k^2)
{\displaystyle {d \Sigma_{0}(k^2)\over{d k^2}}}\over
\big[k^2+\Sigma_{0}^2(k^2)\big]^2}
\right.\nonumber\\
&&\qquad\qquad+\left.{\displaystyle k^6
\left( {d\Sigma_{0}(k^2)\over d k^2}\right)^2
- k^4\Sigma_{0}^2(k^2)\left({d \Sigma_{0}(k^2)\over d k^2 }\right)^2 - k^4
\Sigma_{0}(k^2){d \Sigma_{0}(k^2) \over d k^2 }\over 2
{\big[k^2+\Sigma_{0}^2(k^2)\big]^2}
}\right]
\end{eqnarray}

In Minkowski metrics the propagator (24) has a pole at 
$q^2 = m^2_{\pi} = -2 m \langle\bar{\psi}\psi\rangle_0/F^2$, 
with residue $(\sqrt{2}s_0/F)^2$, where 
$\langle\bar{\psi}\psi\rangle_0=(d(\underline{r})M^3/2\pi^2) c s_0$.
The Adler-Dashen relation (which follows from the symmetries and current
algebra) requires the identification $f^2_{\pi}= F^2$, so that
$q^2 = m^2_{\pi} = -2 m \langle\bar{\psi}\psi\rangle_0/f^2_{\pi}$.   
The rescaling factor $b$ relating the canonical field $\varphi_{\pi}$
(with unit residue at the pole) to the field $\pi$, $\varphi_{\pi}=b\pi$,
is then $b=-f_{\pi}/\sqrt{2}s_0$, as indeed follows from 
current algebra and soft pions theorem (see ref. \cite{bcd}).
Comparison of (25) with the Pagels-Stokar formula (1) leads to
our new formula (2) of the introduction.

To get a numerical insight into the problem we use the dynamical
calculations in ref. \cite{bcd,giulio}, where
overall fits to low energy QCD were made on the basis of two
alternative Ansatz for $\Sigma(k)= s f(k)$: a smooth Ansatz $f(k) =
M^3 /(M^2 + k^2)$ for which the relevant parameters took values
$c=0.554$, $s_0=-4.06$, and a step-function Ansatz 
$f(k) =M [\theta(M^2 -k^2)
+(M^2/k^2) \theta(k^2 - M^2)]$ for which one found $c=0.32$, $s_0=-2.69$. 
Our new  
expression for $f_{\pi}$ has $f^2_{\pi} = f^2_{PS}(1 +\delta^2)$,
where $\delta$ follows from Eq. (25).
 We find $\delta^2 = 0.347$
in the case of the smooth Ansatz, and $\delta^2 = 0.376$ for the
step-function Ansatz.
We want to remark that in the massless limit we consider,
the correction $\delta$ depends only on $s_0$ (or $c$) and the shape of
$f(k)$, but not on $M$. 
In particular, because of the fact that the relevant contribution to the 
chiral symmetry breaking phenomenon comes from relatively short-distance
effects, the corrections will depend mainly on the ultraviolet behaviour
of the self-energy.
That is, the correction does not depend on the fit we
have made to low energy QCD.

It therefore seems that: $(i)$ the corrections are relevant with respect
to the old Pagels-Stokar formula; $(ii)$ the corrections do not
seem to vary in a sensible way when varying the Ansatz for the self energy,
at least within the Ansatz we have used.

It is obvious that the next step one should take is to see whether the 
neglected corrections, which we know must be there, can modify the 
results. However, due to previous experience
 from the study of the pseudoscalar masses \cite{bcd}, 
we would not expect substantial  changes in  the overall
picture of dynamical symmetry breaking.

Finally we may  stress that the new formula (2) we have obtained 
for $f_{\pi}$, within the approximations made, does not require 
additional inputs beyond those already present in the Pagels-Stokar
formula.

\vspace{2cm}

{\sl This work has been carried out within the program Human Capital 
and Mobility (BBW/OFES 95.0200; CHRXCT 94-0579)}


\begin{thebibliography}{99}

\bibitem{pagels} 
H. Pagels and S. Stokar, Phys.Rev. {\bf D20} (1979) 2947.
\bibitem{jackiw}
R. Jackiw and K. Johnson, Phys. Rev. {\bf D8} (1973) 2386.
\bibitem{miranski}
For a review and for additional references:
 V. A. Miransky, ``Dynamical Symmetry Breaking
in Quantum Field Theories", World Scientific, Singapore (1993);
K. Aoki, M. Bando, T. Kugo, M. G. Mitchard and H. Nakatani, Prog. Theor.
 Phys. {\bf 84} (1990) 683;
P. Jain and H. J. Munczek, Phys. Rev {\bf D44} (1991) 1873, 
Phys. Lett. {\bf B282} (1992) 157 ;
T. Kugo and M. G. Mitchard, Phys  Lett. {\bf B282} (1992) 162,
Phys.  Lett. {\bf B286} (1992) 355. 
\bibitem{cjt}
J. M. Cornwall, R. Jackiw and E. Tomboulis, Phys. Rev. {\bf D10} (1974) 2428.
\bibitem{bcd}
A. Barducci, R. Casalbuoni, S. De Curtis, D. Dominici and R. Gatto,
     Phys. Lett. {\bf 147B} (1984) 460, Phys. Rev. {\bf D38} (1988) 238;
R. Casalbuoni, S. De Curtis, D. Dominici and R. Gatto, Phys. Lett.
     {\bf 150B} (1985) 295;
R. Casalbuoni, in {\sl Proceedings of the International Symposium on Composite
Models of Quarks and Leptons}, Tokyo, Japan, 1985, edited by H. Terazawa and 
M. Yasu\'e (Institute for Nuclear Study, University of Tokyo, Tokyo, 1985).
\bibitem{genrev} For general reviews of the field and references see:
S. L. Adler and R. F. Dashen, ``Current Algebra and Applications'', 
Benjamin, New York (1968);
S. B. Treiman, R. Jackiw and D. J. Gross, ``Lectures on Current Algebra
and Applications'', Princeton University Press, Princeton, New Jersey (1972);
V. De Alfaro, S. Fubini, G. Furlan and C. Rossetti, ``Current in Hadron 
Physics'', North Holland, Amsterdam (1973); E. Commins and P. H. Bucksbaum,
``Weak Interactions of Leptons and Quarks'', Cambridge University Press,
Cambridge, England (1983);
H. Georgi, ``Weak Interactions and Modern Particle Theory'', Benjamin-Cummings,
 Menlo Park, California (1984).
\bibitem{giulio}
A. Barducci, R. Casalbuoni, S. De Curtis, R. Gatto and G. Pettini,
     Phys. Rev. {\bf D46} (1992) 2203.
\end{thebibliography}
\end{document}